\documentclass[aps,prl,twocolumn,nofootinbib,superscriptaddress,preprintnumbers,showpacs]{revtex4}
\usepackage{graphicx}

\def\be{\begin{equation}}
\def\ee{\end{equation}}
\def\bea{\begin{eqnarray}}
\def\eea{\end{eqnarray}}
\def\beb{\begin{eqnarray*}}
\def\eeb{\end{eqnarray*}}

\newlength{\myVSpace}% the height of the box
\begin{document}
\def\fmslash{\@ifnextchar[{\fmsl@sh}{\fmsl@sh[0mu]}}
\def\fmsl@sh[#1]#2{%
  \mathchoice
    {\@fmsl@sh\displaystyle{#1}{#2}}%
    {\@fmsl@sh\textstyle{#1}{#2}}%
    {\@fmsl@sh\scriptstyle{#1}{#2}}%
    {\@fmsl@sh\scriptscriptstyle{#1}{#2}}}
\def\@fmsl@sh#1#2#3{\m@th\ooalign{$\hfil#1\mkern#2/\hfil$\crcr$#1#3$}}
\makeatother
\title
{Quarkonia decays into two photons induced by space-time non-commutativity}

\author{B. Meli\'{c}}
\affiliation{Theoretical Physics Division, Rudjer Bo\v skovi\' c Institute, 
Zagreb, Croatia}
\author{K.Passek-Kumeri\v cki}
\affiliation{Theoretical Physics Division, Rudjer Bo\v skovi\' c Institute, 
Zagreb, Croatia}
\author{J.Trampeti\'{c}}
\affiliation{Theoretical Physics Division, Rudjer Bo\v skovi\' c Institute, 
Zagreb, Croatia}
%\affiliation{Theory Division, CERN, CH-1211 Geneva 23, Switzerland}
\affiliation{Max Planck Institut f\"{u}r Physik, 
M\"{u}nchen, Germany}

\date{\today}

\begin{abstract}
%\noindent
In this article we propose  standard model strictly forbidden decay modes,
quarkonia ($\overline{Q}Q_{1^{--}} = J/\psi,\, \Upsilon$) decays into two photons, 
as a possible signature of space-time non-commutativity. 
An experimental discovery of $J/\psi \rightarrow \gamma\gamma$ 
and/or $\Upsilon \rightarrow \gamma\gamma$ processes
would certainly indicate a violation of the 
Landau-Pomeranchuk-Yang theorem and a definitive appearance
of new physics.
\end{abstract}

%\vspace*{1cm}
%{\bf Last updated: jt  23.03.05}
\maketitle
%\end{titlepage}
% ----------------------------------------------------------

%%%%%%%%%%%%%%%%%%%%%%%%%%%%%%%%%%%%%%%%%%%%%%%%%%%%%%%%%%%%%%%%%%
\section{Introduction}
\label{sec:intro}
%%%%%%%%%%%%%%%%%%%%%%%%%%%%%%%%%%%%%%%%%%%%%%%%%%%%%%%%%%%%%%%%%%
%%%

A general feature of gauge theories on non-commutative (NC) space-time is the
appearance of many new interactions which can lead to standard model (SM) 
forbidden processes. 
In this paper we use the
non-commutative standard model (NCSM) to estimate 
decay of heavy quarkonia 
($\overline{Q}Q_{1^{--}} = J/\psi$,\, $\Upsilon$), i.e. quarkonia
annihilation into two photons, which is 
strictly forbidden in the SM 
by angular momentum conservation and Bose statistics,
known as Landau-Pomeranchuk-Yang (LPY) theorem.
Since the violation of the LPY theorem represents in fact 
the violation of Lorentz invariance,
which is intrinsically embedded in non-commutative theories,
such decays can in principle serve as an signature
of space-time non-commutativity. This proposal represents an
attempt to obtain the bound on the NC scale $\Lambda_{\rm NC}$ from
hadronic physics.

A method for implementing non-abelian $SU(N)$ Yang-Mills theories on
non-commutative space-time was proposed in
\cite{Seiberg:1999vs,Madore:2000en,Jurco:2000ja,Jurco:2001rq}. 
In \cite{Calmet:2001na,Behr:2002wx,Melic:2005fm,Melic:2005} this method 
was applied to the
full SM of particle physics resulting
in a minimal non-commutative extension of the SM with the same structure 
group $SU(3)_C \times SU(2)_L \times U(1)_Y$ and with the same
fields and number of coupling
parameters as in the original SM.
It is the only known approach that allows to build models of the
electroweak sector directly based on the structure group $SU(2)_L \times U(1)_Y$
in a non-commutative background.
We call this model NCSM and it represents an effective, anomaly free \cite{Brandt:2003fx}, 
non-commutative field theory. 
Space-time non-commutativity can be
parameterized by the constant antisymmetric matrix 
$\theta_{\mu \nu}$, 
\begin{equation}
 x_{\mu}* x_{\nu}- x_{\nu}* x_{\mu}=i \theta_{\mu\nu}
\nonumber\, ,
\end{equation}
where $*$ denotes the product of non-commutative structure,
while $\theta^{\mu\nu} ={c^{\mu\nu}}/{\Lambda_{NC}^2}$,
with $c^{\mu\nu}$ being dimensionless coefficients 
presumably of order unity.

Violation of Lorentz symmetry is introduced by virtue of 
nonzero $\theta^{\mu \nu}$.
Furthermore, the analysis of discrete symmetry properties of the
NCSM \cite{Aschieri:2002mc} shows that $\theta$
transforms under C, P, T in such a way that it preserves these
symmetries in the action. However, considering $\theta$
as a fixed spectator field, there will be spontaneous
breaking of C, P and/or CP (relative to the spectator).
In the process of interest the C symmetry is violated.

An alternative proposal for the construction of non-commutative
generalizations of the standard model has been put forward in
\cite{Chaichian:2001py}.

Signatures of non-commutativity have been discussed from the point of 
view of collider physics~\cite{Hewett:2000zp,Ohl:2004tn}, including SM forbidden 
$Z \rightarrow \gamma \gamma, gg$ decays \cite{Behr:2002wx,Duplancic:2003hg}, 
neutrino astrophysics \cite{Schupp:2002up}
and neutrino physics \cite{Minkowski:2003jg},  as 
well as low-energy non-accelerator experiments
\cite{Mocioiu:2000ip,Anisimov:2001zc,Carlson:2001sw}. 
Note that the Lorentz violating operators considered in 
\cite{Mocioiu:2000ip} and \cite{Anisimov:2001zc} 
do not appear in the NCSM  
\cite{Madore:2000en,Jurco:2000ja,Jurco:2001rq,Calmet:2001na,Behr:2002wx,Duplancic:2003hg,
Schupp:2002up,Minkowski:2003jg,Melic:2005fm,Melic:2005}
considered in this article.
Furthermore, the rather high bound on non-commutativity obtained in \cite{Carlson:2001sw}
is based on a particular operator contribution appearing in NC QCD that, 
as discussed in \cite{Calmet:2004dn},
is canceled by the contribution of other terms in this model. 
In the NCSM 
\cite{Madore:2000en,Jurco:2000ja,Jurco:2001rq,Calmet:2001na,
Behr:2002wx,Duplancic:2003hg,Schupp:2002up,Minkowski:2003jg,Melic:2005fm,Melic:2005}
the existing bound of $|\theta| \stackrel{<}{\sim} (10\, {\rm TeV})^{-2}$ 
comes from a rather crude model estimate obtained in \cite{Carroll:2001ws}. 
Finally, research of bound state decays 
in the framework of noncommutative QED/QCD 
\cite{Riad:2000vy,Armoni:2000xr,Bonora:2000ga}
were performed by  
computating the lifetimes of ortho and para positronium \cite{Caravati:2002ax} as well as 
the corrections to gluonic decays of heavy quarkonia \cite{Devoto:2004qv}.
%with the set of vertices and Feynman rules from \cite{Riad:2000vy,Armoni:2000xr,Bonora:2000ga}.
For reviews, see \cite{Trampetic:2002eb,Hinchliffe:2002km,Schupp:2004dz}. 

Experimental discovery of the kinematically allowed decays
$J/\psi\rightarrow \gamma\gamma$ and
$\Upsilon\rightarrow \gamma\gamma$, as well as $Z \rightarrow \gamma \gamma$
and $Z \rightarrow gg$,
would certainly prove a violation of the LPY theorem and could serve as a
possible indication/signal for space-time non-commutativity.

In Section 2 we briefly review the ingredients of the NCSM 
relevant to this work.
In Section 3 the amplitudes for the
$\overline{Q}Q_{1^{--}} \to \gamma \gamma$ process are worked out,
while in Section 4 the decay rates are determined.
Section 5 is devoted to the discussion of numerical results
and concluding remarks.

\section{The non-commutative standard model}

The general action of the NCSM is
\begin{equation}
S_{\mbox{\tiny NCSM}}= S_{\mbox{\tiny fermions}} + S_{\mbox{\tiny gauge}} + S_{\mbox{\tiny{Higgs}}} +
S_{\mbox{\tiny{Yukawa}}}
\, ,
\label{eq:Sncsm}
\end{equation}
where for explicit expressions of particular contributions
we refer to \cite{Melic:2005fm}. 

For the simple case of quark QED interactions, 
which is relevant to SM forbidden decays of quarkonia into two photons,
the expansion up to the first order in 
the NC parameter $\theta$ reads
\begin{eqnarray}
S_{\psi, \mbox{\tiny QED}} & = & \int d^4x \, \overline{\widehat \psi} \,
 *  (i \widehat{\slash{\!\!\!\!D}} -m_q)\, \widehat \psi
 \nonumber\\
& = & \int d^4x
\left[ \overline{\psi} \, (i \slash{\!\!\!\!D} -m_q)\, \psi
\right.
\nonumber\\
&& -\left.\frac{1}{4}\,  \overline{\psi}\, A_{\mu \nu} \,
 (i \theta^{\mu \nu \rho} \,  D_{\rho} -m_q \, \theta^{\mu \nu})
  \psi  \right]
\, ,
\label{eq:hatPsiDPsiM}
\end{eqnarray}
where $\theta^{\mu \nu \rho}=
\theta^{\mu \nu} \gamma^{\rho}
+ \theta^{\nu \rho} \gamma^{\mu}
+ \theta^{\rho \mu} \gamma^{\nu}$
and $A_{\mu \nu}$ is the photon field strength tensor.
The kinetic part in (\ref{eq:hatPsiDPsiM}) comes from $S_{\mbox{\tiny fermion}}$, while
the mass contribution originates from  $S_{\mbox{\tiny Yukawa}}$
which, in the case of QED, interactions takes this simple form \cite{Melic:2005fm}.
At the order $\theta$, the electroweak interactions have additional, much more involved 
mass contributions, but only the SM electroweak
quark interactions are relevant to our calculation. 

In the gauge sector of the action (\ref{eq:Sncsm}), 
we have freedom in
the choice of traces in kinetic terms for gauge fields. 
Two different choices were under consideration in \cite{Melic:2005fm} 
producing two different actions in the gauge sector, 
corresponding to the so-called minimal (mNCSM) and 
the non-minimal (nmNCSM) model, respectively.
The matter sector of the action 
is not affected by the change of the gauge part; 
the quark-gauge boson interactions remain the same in both
models.

The mNCSM adopts the following choice for the traces 
in $S^{\rm mNCSM}_{\mathrm{gauge}}$: a sum of three
traces over the $U(1)_Y$, $SU(2)_L$, $SU(3)_C$ sectors
with $Y = \frac{1}{2} \Big({1 \, \phantom{-}0 \atop 0  \,-1}\Big)$
in the definition of $\mbox{Tr}_{1}$ and the
fundamental representation for $SU(2)_L$ and $SU(3)_C$ generators
in $\mbox{Tr}_{2}$ and $\mbox{Tr}_{3}$, respectively.
Up to the first order in the $\theta$ expansion,
the following gauge terms in the mNCSM are obtained:
\begin{eqnarray}
  S^{\rm mNCSM}_{\mathrm{gauge}}&=&-\frac{1}{4} \, \int d^4x 
\, {\cal A}_{\mu \nu} {\cal A}^{ \mu \nu}
\nonumber \\
&&-\frac{1}{2} \, {\rm Tr} \int d^4x \Big[ B_{\mu \nu} B^{\mu \nu}
+ G_{\mu \nu} G^{\mu \nu}
\nonumber \\
& &- 2\, g_S \, \theta^{\mu \nu} \,\left( \frac{1}{4}G_{\mu \nu}
G_{\rho \sigma} - G_{\mu \rho} G_{\nu \sigma}\right) G^{\rho\sigma}\Big] \;.
\nonumber \\
\end{eqnarray}
Here, ${\cal A}_{\mu\nu}$, $B_{\mu\nu}(=B^a_{\mu\nu}T^a_L)$ and $G_{\mu\nu}(=G^a_{\mu\nu}T^a_S)$
denote the $U(1)_Y$, $SU(2)_L$ and $SU(3)_C$ field strengths, respectively.
At order $\theta$ there are no corrections
nor new interactions involving electroweak fields.

One can pick up the other representation of the gauge sector, such that 
the trace is chosen over all particle multiplets
on which covariant derivatives act
and which have different quantum numbers.
In the SM, these are five multiplets for each generation
of fermions and one Higgs multiplet.
New triple neutral gauge boson interactions,
usually forbidden by Lorentz invariance, 
angular moment conservation and the LPY theorem, arise quite naturally
in the framework of nmNCSM model \cite{Behr:2002wx,Duplancic:2003hg}.

In the case of quarkonia decays into two photons, 
the gauge sector in the nmNCSM
gives important contributions through the appearances of novel
triple gauge boson $\gamma\gamma\gamma$ and $Z\gamma\gamma$ couplings. 
Here we give parts relevant to this paper: 
\begin{eqnarray}
{\cal L}^{\rm nmNCSM}_{\gamma\gamma\gamma} &=& 
\frac{e}{4} \sin{2\theta_W} K_{\gamma\gamma\gamma} 
 \nonumber\\
&& \times
\theta^{\rho\sigma} A^{\mu\nu} \left ( A_{\mu\nu} A_{\rho\sigma} 
- 4 A_{\mu\rho} A_{\nu\sigma} \right ),
\nonumber\\
{\cal L}^{\rm nmNCSM}_{Z\gamma\gamma} &=&
\frac{e}{4} \sin{2\theta_W} K_{Z\gamma\gamma} 
  \nonumber\\
&& \times
\theta^{\rho\sigma} \Big[ 2 Z^{\mu\nu} ( 2 A_{\mu\rho} A_{\nu \sigma} - 
A_{\mu\nu} A_{\rho\sigma} ) 
\nonumber\\
&&  
+ 8 Z_{\mu\rho} A^{\mu\nu} A_{\nu\sigma} - 
Z_{\rho\sigma} A_{\mu\nu}A^{\mu\nu} \Big ].
\label{nmNCSM}
\end{eqnarray}
For details of the nmNCSM construction and the allowed values of the constants 
$K_{\gamma\gamma\gamma}$ and $K_{Z\gamma\gamma}$, see \cite{Behr:2002wx,Duplancic:2003hg}.
Parameters in 
the nmNCSM can be restricted by considering GUTs 
on non-commutative space-time \cite{Aschieri:2002mc}.

\section{Amplitudes for $\overline{Q}Q_{1^{--}} \rightarrow \gamma \gamma$ decays}
 
\begin{figure}[h]
\begin{center}
\hspace{-2cm}
\includegraphics[scale=0.43]{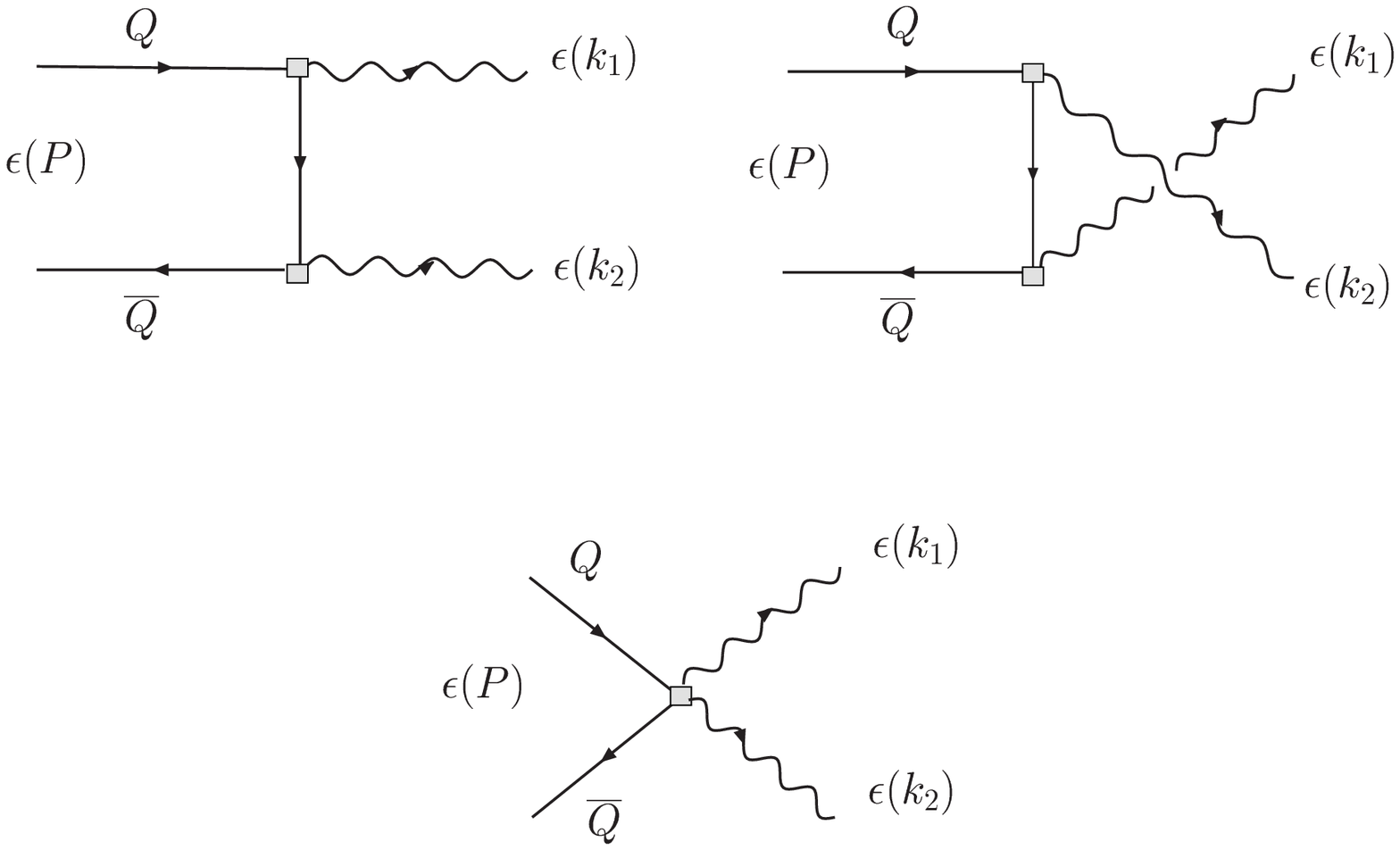}
\end{center}
\caption{Contributions to the 
$\overline{Q}Q_{1^{--}} \rightarrow \gamma \gamma$ amplitude
in the mNCSM, up to the first order in $\theta$.}
\label{f:1}
\end{figure}

\begin{figure}
\begin{center}
\hspace{-1cm}
\includegraphics[scale=0.47,width=0.6\linewidth,height=0.3\linewidth]{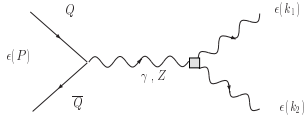}
\end{center}
\caption{Additional contributions to the 
$\overline{Q}Q_{1^{--}} \rightarrow \gamma \gamma$ amplitude in the nmNCSM,
up to the first order in $\theta$.}
\label{f:2}
\end{figure}

In our model, 
for computations of relevant matrix elements, i.e. for
computations of the diagrams from Figures \ref{f:1} and \ref{f:2}, 
we employ Feynman rules 
derived in \cite{Melic:2005fm}.
A helpful property of the action, important for check of calculations, is
its symmetry under ordinary gauge transformations 
in addition to non-commutative ones.
Furthermore, 
in the computations of the diagrams from Figures 
\ref{f:1} and \ref{f:2}, 
applying the following prescription for quarkonia  
to the vacuum transition matrix element of the operator 
$q^{\alpha}_i \overline{q}^{\beta}_j$ ($q =c,b$ and $i,j$ are color indices) 
\begin{eqnarray}
\langle 0 | q^{\alpha}_i \overline{q}^{\beta}_j | \overline{Q}Q_{1^{--}}(P) \rangle  =  
- \frac{|\Psi_{\overline{Q}Q}(0)|}{\sqrt{12 M}} 
\left[(\slash{\!\!\!\!P} + M) \slash{\!\!\!\epsilon}\right]^{\alpha\beta}
\, \delta_{ij},
\end{eqnarray}
we actually hadronize 
free quarks into a quarkonium bound state and calculate 
the amplitude for the quarkonia decay into two photons. 
Here, $|\Psi_{\overline{Q}Q}(0)|$ represents the quarkonia wave function at the origin
\begin{eqnarray}
|\Psi_{\overline{Q}Q}(0)|^2  = \frac{\Gamma(\overline{Q}Q_{1^{--}} \to \ell^+ \ell^-) 
M^2}{ 16 \pi \alpha^2 e_Q^2 }
\, .
\label{wfo}
\end{eqnarray} 

%%%%%%%%%%%%%%%%%%%%%%%%%%%%%%%%%%%%%%%%
%%%%%%%%%%%%%%%%%%%%%%%%%%%%%%%%%%%%%%%%
In the mNCSM, to order $\theta$, the 
diagrams displayed in Figure \ref{f:1} contribute to the amplitude 
${\cal M}_1({\overline{Q}Q_{1^{--}}\rightarrow \gamma\gamma})$ 
given by
\begin{eqnarray}
{\cal M}_1 &=& i\,\pi\,4\sqrt{3\,M}\,\alpha \, e_Q^2 |\Psi_{\overline{Q}Q}(0)| 
\, \epsilon_{\mu}(k_1) \epsilon_{\nu}(k_2) \epsilon_{\rho}(P)
 \nonumber \\ 
& & \times \Big \{
-(k_1 - k_2)^{\rho} \left [ \theta^{\mu\nu} - 2g^{\mu\nu} 
\frac{(k_1 \theta k_2) }{M^2} \right ]   \nonumber \\
& &  + 2 g^{\mu \rho} \left [ (k_1 \theta)^{\nu} 
- 2 k_1^{\nu} \frac{(k_1 \theta k_2) }{M^2} \right ] 
\nonumber \\ 
& & + 2 g^{\nu \rho} \left [ (k_2 \theta)^{\mu}  
+  2 k_2^{\mu} \frac{(k_1 \theta k_2) }{M^2} \right  ]
\Big \} ,
\label{AF1}
\end{eqnarray}
where 
$(k_i \theta)^{\mu} = k_{i\,\nu}\theta^{\nu\mu}$ and 
$k_1\theta k_2 = k_{1\,\mu} \theta^{\mu\nu} k_{2\,\nu}$,
while
$M$ and $P$ are the mass and 
the total momentum of the discussed quarkonium state, respectively.

\begin{figure}
\begin{center}
\includegraphics[scale=.8]{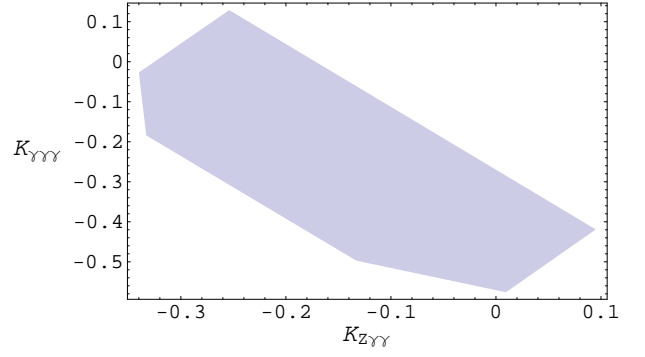}
\end{center}
\caption{The allowed region for the 
$K_{\gamma\gamma\gamma}$ and $K_{Z\gamma\gamma}$ coupling constants \cite{Duplancic:2003hg}.}
\label{f:3}
\end{figure}

Additional diagrams displayed in Figure \ref{f:2}
contribute in the nmNCSM. The corresponding amplitude 
${\cal M}_2({\overline{Q}Q_{1^{--}}\rightarrow \gamma\gamma})$ reads
\begin{eqnarray}
{\cal M}_2 &=& 
-\,i\,\pi \frac{16 \sqrt{3\,M}}{M^{2}} \alpha |\Psi_{\overline{Q}Q}(0)|
\nonumber\\
&&\times \, 
\epsilon_{\mu}(k_1)\epsilon_{\nu}(k_2)\epsilon_{\rho}(P) 
\, \Theta_3((\mu,k_1),(\nu,k_2),(\rho, P)) 
\nonumber\\
&&\times  \left [ e_Q\,\sin2\theta_{W} K_{\gamma\gamma\gamma} + 
\left ( \frac{M}{M_Z} \right )^2 c_V^Q K_{Z\gamma\gamma} \right ]\,,
\label{AF2} 
\end{eqnarray}
where the needed vector couplings of $c$ and $b$ quarks are given as 
\cite{Eidelman:2004wy}
\begin{eqnarray}
c_V^c &=& \frac{1}{2}\left ( 1 - \frac{8}{3} \sin^2\theta_W \right )\,, \nonumber \\
c_V^b &=& -\frac{1}{2}\left ( 1 - \frac{4}{3} \sin^2\theta_W \right )\, ,  
\end{eqnarray}
and $\Theta_3$ is the triple gauge boson function 
defined in \cite{Melic:2005fm,Melic:2005}.
The values of $K_{\gamma\gamma\gamma}$ and $K_{Z\gamma\gamma}$ coupling constants 
evaluated at the $M_Z$ scale
are restricted by the six-sided polygon, 
as shown in Figure \ref{f:3}, taken from \cite{Duplancic:2003hg}.

While only the quark exchange diagrams from Figure \ref{f:1}
contribute to the mNCSM amplitude
${\cal M}^{\rm mNCSM}={\cal M}_1$,
the  nmNCSM amplitude for the quarkonia decay into two photons
amounts to the sum of contributions from the diagrams in 
Figures \ref{f:1} and \ref{f:2}, 
${\cal M}^{\rm nmNCSM}={\cal M}_1+{\cal M}_2$. 

At the end of this section, note that the NCSM free quark amplitude obtained from 
the sum of diagrams in Figure \ref{f:1} is invariant under the 
electromagnetic gauge transformation, as it should be.
Each of the $\gamma$ and $Z$ exchange diagrams from Figure \ref{f:2}
is also electromagnetically gauge invariant, as expected.

\section{Rates of quarkonia decays into two photons}

The amplitude squared for the quarkonia decay derived
in the nmNCSM from the amplitudes (\ref{AF1}) and (\ref{AF2}) is
\begin{eqnarray}
\sum_{\rm pol.}
|{\cal M}^{\rm nmNCSM}(\overline{Q}Q_{1^{--}} \rightarrow \gamma \gamma)|^2 
= \sum_{\rm pol.}|{\cal M}_1 + {\cal M}_2|^2 , 
\nonumber \\
\end{eqnarray}
from which, after the phase space integration, we obtain the decay rate, 
\begin{eqnarray}
\lefteqn{\Gamma^{\rm nmNCSM}(\overline{Q}Q_{1^{--}} \to \gamma \gamma )}
\nonumber \\& =& 
\frac{4 \alpha^2 \,\pi}{3} |\Psi_{\Upsilon}(0)|^2 \,
\frac{M^2}{\Lambda_{\rm NC}^4} \,
\left [7\vec{E}_{\theta}^2 +3 \vec{B}_{\theta}^2  \right ] 
\nonumber \\ & & \times
\left [\frac{e^2_Q}{2} -e_Q \sin 2\theta_W K_{\gamma\gamma\gamma} - 
\left (\frac{M}{M_Z} \right )^2 c_V^Q K_{Z \gamma\gamma }\right ]^2\,.
\nonumber 
\\
\label{pdw}
\end{eqnarray}
The coupling constants appearing in Eq. (\ref{pdw}) are evaluated 
at the $M_Z$ scale \cite{Duplancic:2003hg,Eidelman:2004wy}.
Analogously one obtains the decay rate for mNCSM, 
$\Gamma^{\rm mNCSM}(\overline{Q}Q_{1^{--}} \to \gamma \gamma )$.
It corresponds to 
setting $K_{\gamma\gamma\gamma}\,=\,K_{Z\gamma\gamma}\,=\,0$ in (\ref{pdw}).

In the above computations we have used the following identities:
\begin{eqnarray}
\theta^2 &=&(\theta^2)^{\mu}_{\mu} = \theta_{\mu\nu}\theta^{\nu\mu} 
\nonumber \\
&=& \frac{2}{\Lambda_{\rm NC}^4} \left (\sum_{i=1}^3 (c^{0i})^2 
- \sum_{i,j = 1\,;\; i<j}^3 (c^{ij})^2 \right ) 
\nonumber \\
&\equiv& \frac{2}{\Lambda_{\rm NC}^4}
 \left (\vec{E}_{\theta}^2 - \vec{B}_{\theta}^2 \right )\,.
\end{eqnarray}

To maximize the rates (\ref{pdw}) we can assume that the dimensionless
quantities $\vec{E}_{\theta}^2$ and $\vec{B}_{\theta}^2$ are of order one
\footnote{The parameterization of the $\theta ^{\mu\nu}$
matrix elements used here is taken from \cite{Hewett:2000zp}.}. 
Normalizing the obtained decay rate 
to the decay of $\overline{Q}Q_{1^{--}}$ into lepton pairs,
by using (\ref{wfo}) we find
\begin{eqnarray}
\frac{\Gamma^{\rm nmNCSM}(\overline{Q}Q_{1^{--}} \to \gamma \gamma )}
{\Gamma(\overline{Q}Q_{1^{--}} \to \ell^+ \ell^-)}
&=&
\frac{5}{24}\,e^2_Q\,\left (\frac{M}{\Lambda_{\rm NC}} \right )^4
\nonumber \\ & & \times
\left[ 1 -\frac{2}{e_Q} \sin 2\theta_W K_{\gamma\gamma\gamma} 
\right. \nonumber \\ & &
\left.
- 
\frac{2}{e^2_Q}\left (\frac{M}{M_Z} \right )^2 c_V^Q K_{Z \gamma\gamma }\right]^2 \,. 
\nonumber \\
\label{pdwr}
\end{eqnarray}

Hence, in the case of mNCSM couplings, 
we obtain the following ratios
for $\Upsilon $ and $J/\psi$ decays:
\begin{eqnarray}
\frac{\Gamma^{\rm mNCSM}(\Upsilon \to \gamma \gamma )}{\Gamma(\Upsilon \to \ell^+ \ell^-)} =
\frac{5}{216}\left (\frac{M_{\Upsilon}}{\Lambda_{\rm NC}} \right )^4 \,
\end{eqnarray}
and 
\begin{eqnarray}
\frac{\Gamma^{\rm mNCSM}(J/\psi \to \gamma \gamma )}{\Gamma(J/\psi \to \ell^+ \ell^-)} =
\frac{5}{54}\left (\frac{M_{J/\psi}}{\Lambda_{\rm NC}} \right )^4\,.
\end{eqnarray}
Note here that choosing $\vec{E}_{\theta}^2 = 0$ and $\vec{B}_{\theta}^2 \simeq 1$, 
as it is favored by the string theory, would produce result (\ref{pdwr}) multiplied by 3/10,
which does not change the final conclusion in a serious way. 

The range of the scale of non-commutativity, 
$1 \, \geq \Lambda_{\rm NC}/{\rm TeV} \geq 0.25$, 
was choosen because it produces experimentally reachable 
quarkonia decay to two photons rates. 
Since the rate (\ref{pdwr}) depends on $1/{\Lambda^4_{\rm NC}}$ it is quite clear that 
any larger $\Lambda_{\rm NC}$ would dramatically decrease 
possibility to see the signal for  
the non-commutativity of space-time via quarkonia to two photons decay at present and near-future experiments. 

The choosen range of the scale of non-commutativity then gives
\begin{eqnarray}
2\times 10^{-10}\,\stackrel{<}{\sim}
\frac{\Gamma^{\rm mNCSM}(\Upsilon \to \gamma \gamma )}{\Gamma(\Upsilon \to \ell^+ \ell^-)} 
 \stackrel{<}{\sim} \, 5\times 10^{-8}\,
\end{eqnarray}
and 
\begin{eqnarray}
9\times 10^{-12}\,\stackrel{<}{\sim}
\frac{\Gamma^{\rm mNCSM}(J/\psi \to \gamma \gamma )}{\Gamma(J/\psi \to \ell^+ \ell^-)}
 \stackrel{<}{\sim} \, 2\times 10^{-9}\,.
\end{eqnarray}

From the experiment one has 
$\Gamma^{\rm exp.}(\Upsilon(1S) \to e^+ e^-)=(1.314 \pm 0.029)\, {\rm keV}$ and 
$\Gamma^{\rm exp.}_{\rm tot}(\Upsilon(1S))=(53.0 \pm 1.5) \, {\rm keV}$ \cite{Brambilla:2004wf}, 
which then leads to
\begin{eqnarray}
5\times 10^{-12}\,\stackrel{<}{\sim} BR^{\rm mNCSM}(\Upsilon(1S) \to \gamma \gamma) \stackrel{<}{\sim} 10^{-9}.
\label{mBR1}
\end{eqnarray}

For the $J/\psi$ case,  
$\Gamma^{\rm exp.}(J/\psi \to e^+ e^-)=(5.4 \pm 0.15 \pm 0.07)\, {\rm keV}$ and 
$\Gamma^{\rm exp.}_{\rm tot}(J/\psi)=(91.0 \pm 3.2) \, {\rm keV}$
\cite{Brambilla:2004wf}, 
which, with $\Lambda_{\rm NC}$ in the above range, gives 
the following range for the $J/\psi$ branching ratio:
\begin{eqnarray}
5\times 10^{-13}\,\stackrel{<}{\sim} BR^{\rm mNCSM}(J/\psi \to \gamma \gamma)  \stackrel{<}{\sim}\,  10^{-10}.
\label{mBR2}
\end{eqnarray}

For the nmNCSM, 
with the choice of the triple gauge boson couplings $K_{\gamma\gamma\gamma}=-0.576$ 
and $K_{Z\gamma\gamma}=0.010$ 
the rates for the $\Upsilon \to \gamma \gamma$ 
and $J/\psi \to \gamma \gamma$ decays reach maximal values. 
In the same
range of the scale of non-commutativity as above, 
$1\,\geq \Lambda_{\rm NC}/{\rm TeV} \geq 0.25$, we then have
\begin{eqnarray}
7\times 10^{-10}\,\stackrel{<}{\sim} 
\frac{\Gamma^{\rm nmNCSM}(\Upsilon \to \gamma \gamma )}{\Gamma(\Upsilon \to \ell^+ \ell^-)} 
 \stackrel{<}{\sim} \,2\times 10^{-7}\,
\end{eqnarray}
and
\begin{eqnarray}
5\times 10^{-11}\,\stackrel{<}{\sim} 
\frac{\Gamma^{\rm nmNCSM}(J/\psi \to \gamma \gamma )}{\Gamma(J/\psi \to \ell^+ \ell^-)} 
 \stackrel{<}{\sim} \, 10^{-8}\,.
\end{eqnarray}
Using the aforementioned experimental values for $\Upsilon$ and $J/\psi$ we obtain
the following ranges for relevant branching ratios:
\begin{eqnarray}
2\times 10^{-11}\,\stackrel{<}{\sim} BR^{\rm nmNCSM}(\Upsilon \to \gamma \gamma) \stackrel{<}{\sim} \,4\times 10^{-9}\,,
\label{nmBR1}\\
3\times 10^{-12}\,\stackrel{<}{\sim} BR^{\rm nmNCSM}(J/\psi \to \gamma \gamma) \stackrel{<}{\sim} \,8\times 10^{-10} \,.
\label{nmBR2}
\end{eqnarray}

\section{Discussion and conclusion}

\begin{figure}[ht]
\begin{center}
\includegraphics[scale=0.42]{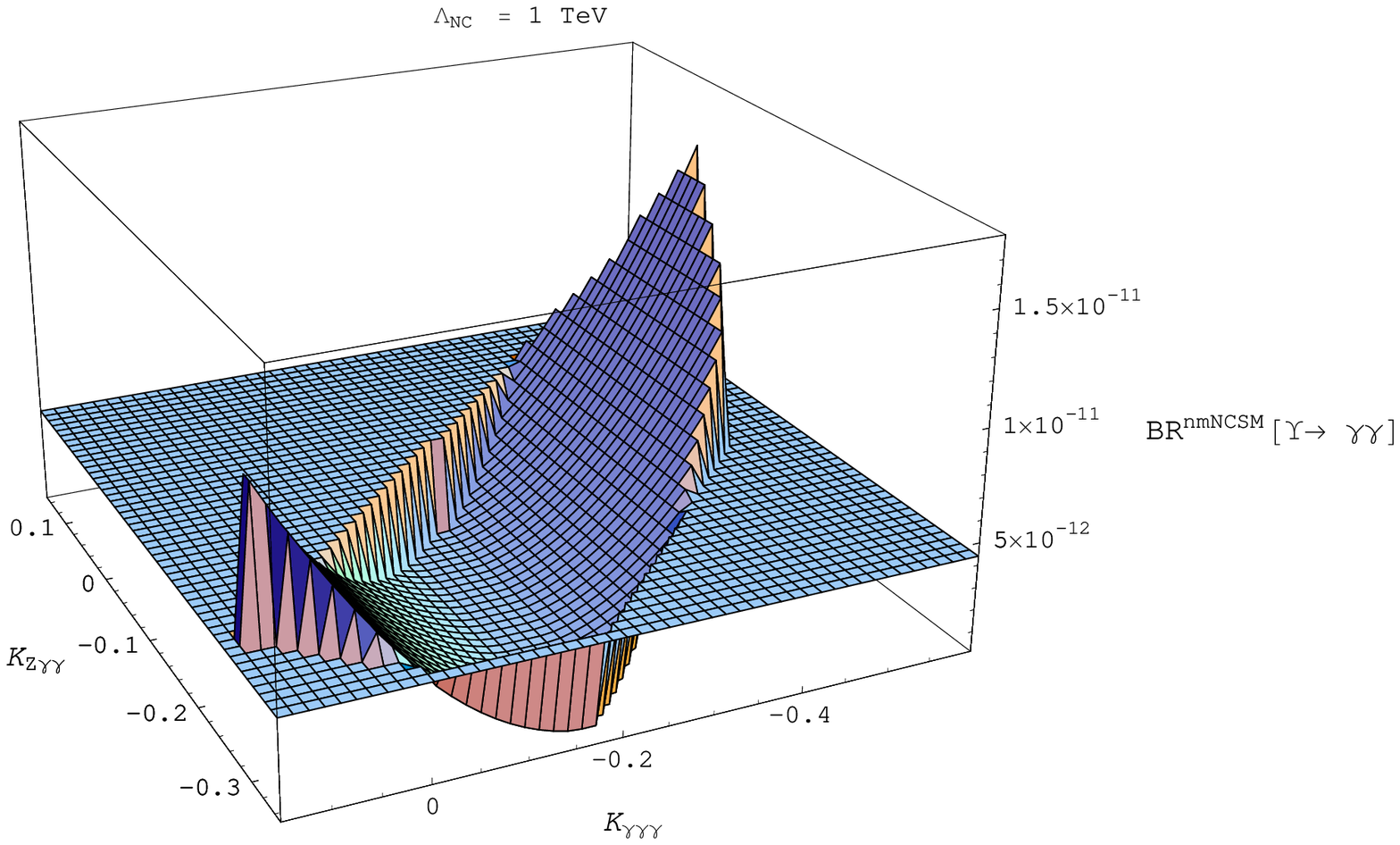}
\end{center}
\caption{Branching ratio $BR^{\rm nmNCSM}(\Upsilon \rightarrow \gamma \gamma)$
as a function of $K_{\gamma\gamma\gamma}$ and $K_{Z\gamma\gamma}$ 
coupling constants, at the scale of non-commutativity $\Lambda_{\rm NC}=1$ TeV. 
The horizontal plane at the value of $4.7\times 10^{-12}$
indicates the 
$BR^{\rm mNCSM}(\Upsilon \rightarrow \gamma \gamma)$,
which one obtains by setting $K_{\gamma\gamma\gamma}=K_{Z\gamma\gamma}=0$ in (\protect{\ref{pdw}}).}
\label{f:4}
\end{figure}
\begin{figure}[t]
\begin{center}
\includegraphics[scale=0.46]{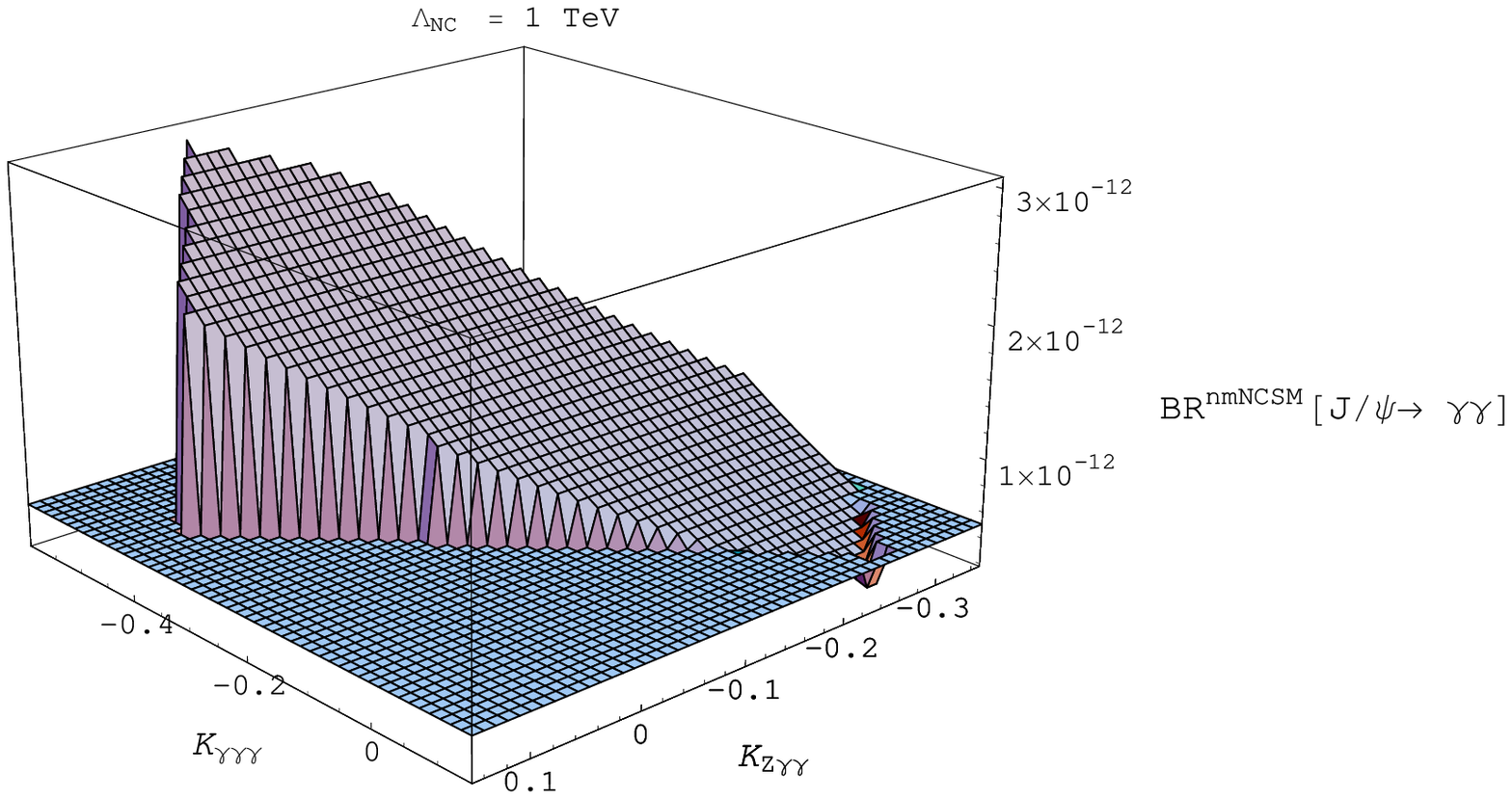}
\end{center}
\caption{Branching ratio $BR^{\rm nmNCSM}(J/\psi \rightarrow \gamma \gamma)$
as a function of $K_{\gamma\gamma\gamma}$ and $K_{Z\gamma\gamma}$ 
coupling constants,  at the scale of non-commutativity $\Lambda_{\rm NC}=1$ TeV.
The horizontal plane at the value of $5.1\times 10^{-13}$
indicates the 
$BR^{\rm mNCSM}(J/\psi \rightarrow \gamma \gamma)$,
which one obtains by setting $K_{\gamma\gamma\gamma}=K_{Z\gamma\gamma}=0$  in (\protect{\ref{pdw}}).}
\label{f:5}
\end{figure}
In this paper  
we have considered decays of two
quarkonia states: $\overline{Q}Q_{1^{--}}=J/\psi,\; \Upsilon(1S)$
into two photons which violate the LPY theorem. 

Theoretically, in the $\Upsilon$ case, the addition of triple neutral gauge boson couplings 
via a photon and a Z-boson exchange diagram in Figure \ref{f:2},
contributes to the nmNCSM amplitude
both constructively and destructively, 
depending on the specific values of the 
$K_{\gamma\gamma\gamma}$ and $K_{Z\gamma\gamma}$ coupling constants
from the area in Figure \ref{f:3}. 
As a consequence, the branching ratio for the $\Upsilon \rightarrow \gamma\gamma$ decay 
increases or decreases with respect to the values obtained in the mNCSM. 
This is illustrated in Figure \ref{f:4}
which shows the branching ratio 
$BR^{\mbox{\tiny nmNCSM}}(\Upsilon \rightarrow \gamma \gamma)$ as
a function of the triple neutral gauge boson constants
$K_{\gamma\gamma\gamma}$ and $K_{Z\gamma\gamma}$.
Since the constructive -- destructive contributions are spread approximatively equally 
within the allowed range of the $K_{\gamma\gamma\gamma}$, $K_{Z\gamma\gamma}$
constants, we conclude that additional uncertainty in the prediction of
the $\Upsilon (1S) \rightarrow \gamma \gamma$ branching ratio, due to the NCSM model building,
makes the $\Upsilon (1S) \rightarrow \gamma \gamma$ decay a less favorite
candidate for distinguishing the mNCSM from the nmNCSM. 

However, for the $J/\psi$ decay, because of the positive charm quark charge, 
the addition of the
photon and Z-boson exchange diagrams (Figure \ref{f:2})
to the quark exchange diagrams in Figure \ref{f:1} contributes constructively 
to the dominant part of the $K_{\gamma\gamma\gamma}\,-\,K_{Z\gamma\gamma}$ area,
producing rates larger by 
a factor of $\simeq 5$ with respect to those gained in the mNCSM.
Destructive contributions are small, covering about a few \% of the 
$K_{\gamma\gamma\gamma}\,-\,K_{Z\gamma\gamma}$ area.
This is illustrated in Figure \ref{f:5} showing that the nmNCSM contributions to the 
$J/\psi \rightarrow \gamma \gamma$ decay are almost always constructive,
thus effectively enhancing its branching ratio.

Experimentally \cite{sk,Brambilla:2004wf},  
concerning the $\Upsilon(1S)$ decay, at CLEO-III there is the largest  
sample of
$\Upsilon(1S)$ resonances produced in $e^+e^-$ collisions, about 21 million. 
However, it will be very difficult to observe the 
$\Upsilon(1S) \to \gamma \gamma$ decay because of the larger
QED background from the non-resonant $e^+e^- \rightarrow \gamma \gamma$ process.
The resonant cross-section for $e^+e^- \rightarrow \Upsilon (1S)$ is
of the same order of magnitude as the background
cross-section. Thus it seems that the detection of
$BR(\Upsilon (1S) \rightarrow \gamma \gamma)$ below $10^{-3}$ will be hopeless
with the present data.
Considering the $J/\psi \to \gamma \gamma$ decay, there are much better chances since
the resonant cross-section is much higher.
The existing limit, which comes from a very old
experiment $BR(J/\psi \rightarrow \gamma \gamma)< 5 \times 10^{-4}$
\cite{Eidelman:2004wy}, 
can be improved but probably the NC limits given in 
(\ref{mBR2}) and (\ref{nmBR2}) are unreachable today. 
Unfortunately, the above experimental limits are too weak 
to set any reliable bound on
the non-commutative scale from our model estimate. 

Finally, note that quarkonia and the Z-boson to two photon decay processes
are related in the nmNCSM, via the $Z\gamma\gamma$ interaction \cite{Behr:2002wx} 
of the strength
$2e\sin2\theta_WK_{Z\gamma\gamma}$ determined in \cite{Duplancic:2003hg}. 
We have found 
that the $\overline{Q}Q_{1^{--}} \rightarrow \gamma \gamma$ decay rates
become maximal for the values $K_{\gamma\gamma\gamma}=-0.576$ and $K_{Z\gamma\gamma}=0.01$. 
The same value $K_{Z\gamma\gamma}=0.01$,  
produces the minimal value of $BR(Z \to \gamma\gamma)$ 
via Eq. (17) from \cite{Behr:2002wx}
and Table 1 from \cite{Duplancic:2003hg}. 
On the other hand, the value of $K_{Z\gamma\gamma}=-0.34$, for any 
$-0.03>K_{\gamma\gamma\gamma}>-0.19$
(see Table 1 in \cite{Duplancic:2003hg}), 
maximizes the $Z \to \gamma\gamma$
decay rate and, at the same time, minimizes 
the $\overline{Q}Q_{1^{--}} \rightarrow \gamma \gamma$ branching ratios
(see Figures \ref{f:4} and \ref{f:5}). 
The combination of all three decays would certainly narrow the parameter
space of unknown constants of our model, like $\theta^{\mu\nu}$, $K_{Z\gamma\gamma}$, etc.

In conclusion, if the future experiments measure any of the $Z \to \gamma\gamma$, 
$J/\psi \rightarrow \gamma \gamma$ and $\Upsilon (1S) \rightarrow \gamma \gamma$
branching ratios, 
the appearances of the physics beyond the SM would 
then be strongly indicated. 
We hope that the importance of a possible
discovery of space-time non-commutativity will convince
experimentalists to look for SM forbidden decays in hadronic physics.
%A good reason for this is that the sensitivity to 
%the non-commutative parameter $\theta^{\mu\nu}$ could be 
%in the  range of 
%the new generation 
%of linear colliders with a c.m.e. around TeV scale. 
%of heavy quark experiments (BES III, CLEO-c).

\subsection*{Acknowledgment}

We want to thank to X. Calmet, G. Duplan\v ci\' c, P. Schupp, 
T. Skwarnicki and J. Wess for fruitful discussions. 
This work was supported by the
Ministry of Science, Education and Sport of the Republic of Croatia
under Contract No. 0098002.

\end{document}